\documentclass[aps,prd,twocolumn,groupedaddress]{revtex4}
\usepackage{graphicx}
\usepackage{dcolumn}
\usepackage{bm}
\usepackage{amsmath}
\usepackage{epstopdf}
\usepackage{amsfonts}
\usepackage{amssymb}%

\usepackage{natbib}
\usepackage[colorlinks=true,citecolor=blue,urlcolor=magenta,breaklinks]{hyperref}

\providecommand{\U}[1]{\protect\rule{.1in}{.1in}}

\begin{document}

\title{Constraining the parameter space of branon dark matter \\
using white dwarf stars}

\author{Grigorios Panotopoulos and Il\'\i dio Lopes}
\email{grigorios.panotopoulos@tecnico.ulisboa.pt, ilidio.lopes@tecnico.ulisboa.pt}
\affiliation{CENTRA, Instituto Superior T{\'e}cnico,\\ Universidade de Lisboa,
Av. Rovisco Pais 1, Lisboa, Portugal}

\date{\today}

\begin{abstract}
In the present work we study the branon dark matter particles impact on compact objects, and we provide the first
constraints of the parameter space using white dwarf stars. The branon dark matter model is
characterized by two free parameters, namely the branon mass particle M and the brane tension factor $f$. The latter determines
the strength of the interaction of branon dark matter particles with baryons.
By considering a typical white dwarf star we were able to obtain constraints on branon dark matter
and compare with current limits obtained by direct detection searches and dark matter abundance.
In particular our results show that i) for heavy branons with a mass $M > 10 GeV$ white dwarfs fail to provide us with bounds better than current limits 
from DM direct detection searches, and ii) for light branons in the mass range $2 keV < M < 1 GeV$, which cannot be probed neither with current dark 
matter experiments nor with the next generation of detectors, the dark matter abundance constrain determines
$f$ as a function of $M$ in the range $0.1 GeV < M < 1 GeV$ for the branon mass and $1 GeV < f < 5 GeV$ for the brane tension factor.
Furthermore, our findings indicate that the limits from white dwarfs are not stronger than the dark matter abundance constrain.
\end{abstract}

\pacs{95.35.+d, 95.30.Cq}
\maketitle

\section{Introduction}

All the available astrophysical and cosmological observational data coming from many different sides
show that the non-relativistic matter in the universe is dominated by so-called Dark Matter (DM). This term
was introduced in 1933 by Zwicky \cite{zwicky} studying clusters of galaxies, and much later in 1970 Rubin and
Ford \cite{rubin} with optical studies of M31 made the case for DM in galaxies. Although as of today there are
many candidates \cite{taoso}, the origin and nature of DM still remains a mystery, and comprises one of the
biggest challenges in modern theoretical cosmology. For a review on dark matter see e.g. \cite{munoz}. A popular class of
DM candidates is the so-called Weakly Interacting Massive Particles (WIMPs), that are thermal relics from the
Big-Bang. Initially the temperature of the Universe was high enough to maintain the DM particle in equilibrium with
the rest of the particles. However, as the Universe expands and cools down at some point the annihilation rate of WIMPs
$\Gamma=n \langle \sigma v \rangle$, with $n$ being the WIMP number density and $\langle \sigma v \rangle$ being the thermal average of
the WIMP annihilation cross section, drops below the Hubble parameter $H$, which measures the expansion rate of the universe. When this happens 
WIMPs can no longer annihilate, and their current abundance is the same ever since.
It turns out that their today's relic density is given by \cite{Jungman:1995df}
\begin{equation}
\Omega_{DM} h^2 = \frac{3 \times 10^{-27} cm^3/s}{\langle \sigma v \rangle},
\end{equation}
where $h$ is related to the Hubble constant $H_0=100 \: h (km s^{-1})/(Mpc)$. If the DM particle has only weak interactions
(besides gravity of course), the WIMPs annihilation cross section typically has a value $\langle \sigma v \rangle = 3 \times 10^{-26} cm^3/s$ \cite{munoz}, and thus reproduces the observed DM abundance $\Omega_c h^2=0.1198 \pm 0.0015$ \cite{planck2015}.
In this work we will focus our study in a special class of WIMPs known as branon dark matter particles.

Superstring theory \cite{strings1,strings2} is so far the only consistent theory of quantum gravity, and since it claims
to give us a fundamental description of Nature, it would be interesting to see what kind of phenomenology and cosmology
it predicts. A well-studied case is the brane-world idea, according to which our four-dimensional
world and the Standard Model (SM) of particle physics are confined to live on a three-dimensional brane, while
gravity lives in the higher-dimensional bulk. Since the higher-dimensional Planck mass $M_D$ is different
than the usual four-dimensional one $M_p$, the brane concept has been used to address the hierarchy problem
of particle physics, first in a flat ($D=4+d$) spacetime with four large dimensions and $d$ small compact dimensions
\cite{ADD1,ADD2}, and later refined by Randall and Sundrum \cite{RS1,RS2}. For a review on brane cosmology see e.g. \cite{branes}.
Since gravity lives in the bulk, the gravitational potential exhibits higher-dimensional behavior $V(r) \sim 1/r^{d+1}$
at small distances $r \ll R$, with $R$ being the size of the extra dimensions, while at large distances $r \gg R$ the Newton's law $V(r) \sim 1/r$ is recovered, and the usual four-dimensional Planck mass $M_p$ is related to the higher-dimensional one $M_D$ as follows \cite{branes}
\begin{equation}
M_p^2 = M_D^{d+2} R^d.
\end{equation}

Since the notion of a completely rigid body is incompatible with Einstein's relativity, the brane fluctuations
must be taken into account. These fluctuations are parameterized by some $\pi$
fields called the branons. They are scalar fields, and as the translational
invariance is explicitly broken, branons can be understood as massive pseudo Nambu-Goldstone bosons. It has been shown in \cite{branons} that if the brane is
flexible, $f \ll M_D$, where the brane tension is written $V=f^4$, the only relevant degrees of freedom on the brane
are the SM fields and the branons, namely the system is described by the action \cite{branons}
\begin{equation}
S=\int d^4x \sqrt{-g} (-f^4 + \mathcal{L}_{SM} + \mathcal{L}_{Br} + \mathcal{L}_{int}),
\end{equation}
where $\mathcal{L}_{SM}$ is the Lagrangian corresponding to the SM of particle physics, $\mathcal{L}_{Br}$
is the branon Lagrangian
\begin{equation}
\mathcal{L}_{Br} = \frac{1}{2} \left( \delta_{\alpha \beta} \partial_\mu \pi^\alpha \: \partial^\mu \pi^\beta - M_{\alpha \beta}^2 \pi^\alpha \pi^\beta \right),
\end{equation}
with $M_{\alpha \beta}$ being the branon mass matrix and the indices $\alpha, \beta$ take values from one to $d$, and $\mathcal{L}_{int}$ is the lagrangian interaction between the branons and the SM fields \cite{branons}
\begin{equation}
\mathcal{L}_{int} = \frac{1}{8 f^4} \left( 4 \delta_{\alpha \beta} \partial_\mu \pi^\alpha \partial_\nu \pi^\beta - M_{\alpha \beta}^2 \pi^\alpha \pi^\beta g_{\mu \nu} \right) T_{SM}^{\mu \nu},
\end{equation}
with $T_{SM}^{\mu \nu}$ being the SM stress-energy tensor. In the rest of the article we shall assume for simplicity that all branons have the same mass M. From the structure of the Lagrangian interaction it is clear that
in interaction vertices branons always appear in pairs and thus they are stable. Since these new particles are massive and weakly coupled, branons are natural dark matter candidates. Indeed branons have been shown to be excellent dark matter candidates \cite{branons2} satisfying the DM constraint $\Omega_c h^2=0.1198 \pm 0.0015$ \cite{planck2015}, while constraints on the {$M-f$} parameter space from astrophysics, cosmology and colliders have been studied in \cite{branons3,branons5}.

To shine some light into the nature of dark matter several earth based experiments have been designed.
In these experiments an effort is made to observe the nucleus recoil after a dark matter particle scatters
off the material of the detector. These direct detection experiments have put limits on the nucleon-dark matter
candidate cross section for a given mass of the dark matter particle \cite{detection1,detection1b,detection2}, while the prospects of branon direct detection have been presented in \cite{branons4}. During the last 15 years or so
observational data from astrophysical objects, such as the Sun~\cite{ilidio1,ilidio2,ilidio3}, solar-like stars~\cite{ilidio4,ilidio5,ilidio6}, white dwarfs and neutron stars~\cite{kouvaris0, kouvaris1, kouvaris2}, have
been employed to offer us complementary bounds on the WIMP-nucleon cross section, see e.g.~\cite{ilidio0} and references therein.

In the present article we use white dwarf stars to constrain the parameter space of the branon
dark matter. Our work is organized as follows: after this introduction, we present the theoretical framework
in section two, and we constrain the branon parameter space in the third section. Finally we conclude in section four.
We work in units in which the speed
of light in vacuum $c$, the Boltzmann constant $k_B$ and the reduced Planck mass $\hbar$ are set to unity, $c=k_B=\hbar=1$.
In these units all dimensionful quantities are measured in GeV, and we make use of the conversion rules
$1 m = 5.068 \times 10^{15} GeV^{-1}$, $1 kg = 5.610 \times 10^{26} GeV$ and $1 K = 8.617 \times 10^{-14} GeV$ \cite{guth}.

\section{Theoretical framework}

\subsection{White dwarfs}

White dwarf (WD) stars are old compact objects that mark the final evolutionary stage of the vast majority of the stars \cite{ref1, ref2}. Indeed more than 95, perhaps up to 98 per cent of all stars will die as white dwarfs \cite{ref3}. They were discovered in 1914 when H. Russell noticed that the star now known as 40 Eridani B was located well below the main sequence on the Hertzsprung-Russell diagram. About 80 per cent of WD show hydrogen atmosphere (DA type), while 20 per cent show helium atmosphere (DB type) \cite{ref4}. The low-mass white dwarfs are expected to harbor He cores, while the average mass white dwarfs most likely contain Carbon/Oxygen cores \cite{ref1}. Since there are no thermonuclear reactions for WD, these objects are cooling down by eradiating. Along the cooling track, there are basically three classes of white stars, namely DAV stars with an effective temperature around $T_{eff} \sim 12 \times 10^3 K$, DBV stars with an effective temperature around $T_{eff} \sim 25 \times 10^3 K$, and DOV white dwarfs with an effective temperature around $T_{eff} \sim 100 \times 10^3 K$ \cite{ref2}.
Here we shall consider a typical white dwarf star of the DBV type with mass $M_{\star}  \sim M_{\odot}  = 2 \times 10^{30} kg$, radius
$R_{\star}  \sim R_{earth} \simeq 0.01 R_{\odot}  \simeq 7 \times 10^3  km$, temperature $T \sim 25 \times 10^3  K$, matter density $\rho \sim 10^{10} kg/m^3$ and pressure $P \sim 10^{23} N/m^2$, where $M_{\odot}$ and $R_{\odot}$ are the solar mass and solar radius respectively. For simplicity we shall assume that 
its core is made exclusively of a single chemical element, such as $O^{16}$, which we show to provide the stringent bound on branon DM.

\subsection{The branon-nucleon cross section}

The branons once trapped inside the star interact with the nuclei and eventually thermalize, and since they are
non-relativistic they are described by the Maxwell-Boltzmann distribution \cite{ilidio1,ilidio1a,ilidio5}. If a large 
number of them is accreted during the lifetime of a white dwarf, they may collapse
and form a small black hole (BH) inside the star that eventually destroy the compact object \cite{catastrofe}. Therefore, the existence of old white dwarfs can impose constraints on the properties of branons. It thus becomes
clear that the most important quantity for the discussion is the branon-nucleon cross section $\sigma_n$, which from the theory side is determined by the two free parameters of the model, namely the branon mass M and the brane tension factor $f$, while from
the experiment side is constrained from direct detection searches, roughly $\sigma_n < 10^{-44} cm^2$ \cite{detection1,detection1b,detection2}.

The Feynman rules for the interaction vertices between branons and the standard model fields have been derived in
\cite{branons1}. Neglecting the difference between neutron and proton, the branon-nucleon scattering cross section
is given by \cite{branons2,branons4}
\begin{equation}
\sigma_n = \frac{9 M^2 m_n^2 \mu^2}{64 \pi f^8},
\end{equation}
where $m_n$ is the nucleon mass, taken to be equal to the mass of the proton  $m_p \simeq 1 GeV$, and $\mu=M m_n/(M+m_n)$ is the reduced mass of the branon-nucleon system. Let us comment that in general the total cross section has a spin-dependent and a coherent (spin-independent) contribution. Depending on the model and on the circumstances, one of the two contributions can dominate over the other, or it is absent all together.
Just to mention a couple of typical examples, during scattering off nuclei with an even number
of nucleons, since cancellations occur between nucleon pairs, the spin-dependent contribution becomes negligible compared
to the coherent one. Or, in models with a scalar dark matter particle the scattering cross section has only spin-independent
contribution \cite{classification}. The same holds for the branon dark matter case \cite{branons4}. So the above
cross section is purely spin-independent (coherent) with no spin-dependent contribution. Now if a nucleus consists
of A nucleons in total, the branon-nucleus scattering cross section is given by \cite{branons4}
\begin{equation}
\sigma = A^2 \sigma_n.
\end{equation}

\begin{figure}[ht!]
	\centering
	\includegraphics[width=\linewidth]{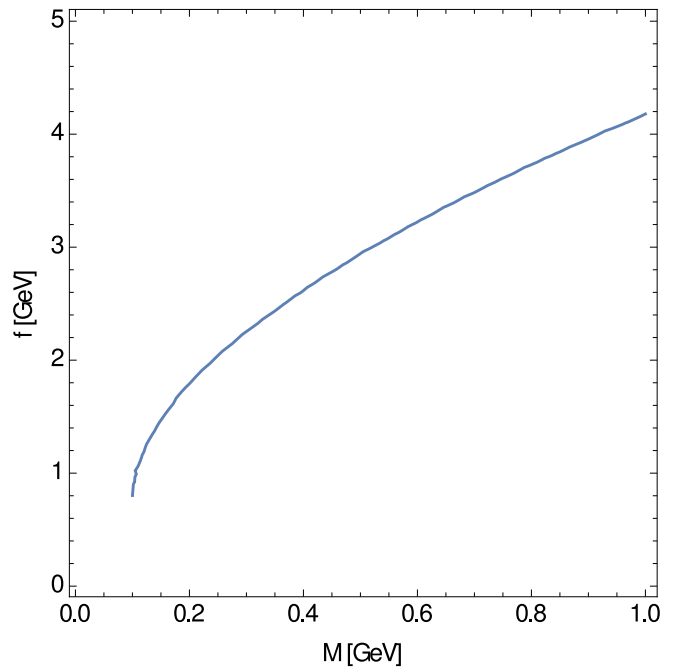}
	\caption{The brane tension factor $f$ (in GeV) as a function of the branon mass M (also in GeV) requiring that $\langle \sigma v \rangle \simeq 3 \times 10^{-26} cm^3/s$.}
\end{figure}

\begin{figure}[ht!]
	\centering
	\includegraphics[width=\linewidth]{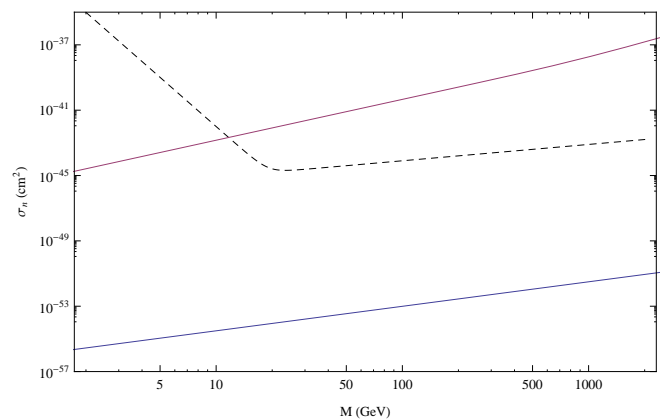}
	\caption{Branon-nucleon cross section in $cm^2$ versus branon mass M in GeV for $A=16$. Shown are the thermalization condition
	(lower solid curve), the condition for the formation of a BEC (upper solid curve), while the dashed curve corresponds to the existing limits from 
	direct detection experiments~\cite{detection1,detection1b,detection2}.}
\end{figure}

\begin{figure}[ht!]
	\centering
	\includegraphics[width=\linewidth]{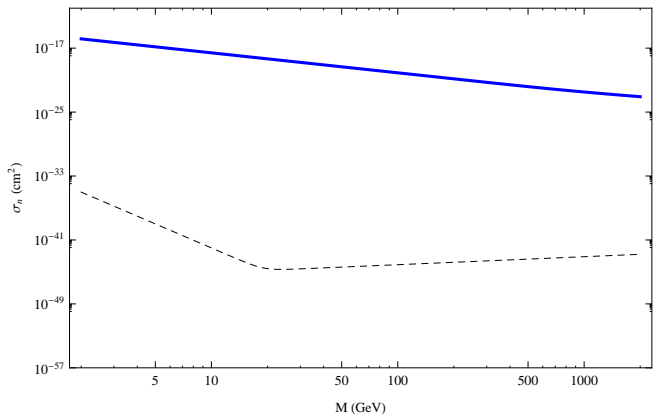}
	\caption{Branon-nucleon cross section in $cm^2$ versus branon mass M in GeV for $A=16$ and heavy branons. The solid curve corresponds to the black 
	hole formation, while the dashed curve corresponds to the existing limits from direct detection experiments~\cite{detection1,detection1b,detection2}.}
\end{figure}

\begin{figure}[ht!]
	\centering
	\includegraphics[width=\linewidth]{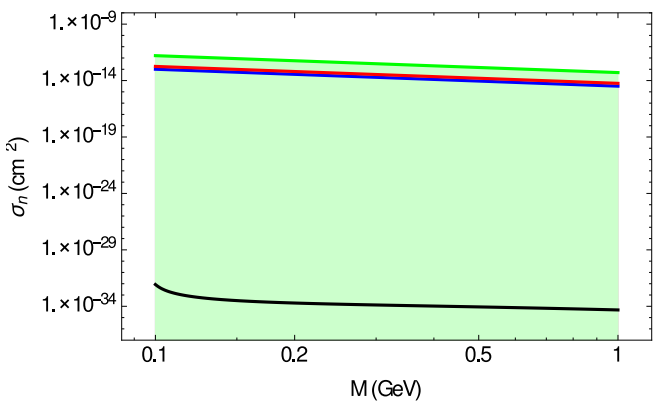}
	\caption{Same as Fig. 3, but for light branons with a mass range $2 keV < M < 1 GeV$, and for $A=4,12,16$ from top to bottom. There is no known experimental constraint in this part of the parameter space. The black curve corresponds to the branon-nucleon cross section computed in the framework of the model (eq. (6)) taking into account the condition $\langle \sigma v \rangle \simeq 3 \times 10^{-26} cm^3/s$.}
\end{figure}

\subsection{The conditions required for the BH formation}

To see if there is enough branon accretion to collapse and form a black hole inside the star, we need to compute the accretion rate \cite{kouvaris1}
\begin{equation}
F =\frac{8 \pi^2 }{3}  \frac{\rho_{dm} }{M}  G M_{\star}  R_{\star}
\left( \frac{3}{2 \pi v^2}\right)^{3/2}  v^2 \left(1-e^{-3 \frac{E_0}{v^2}}\right) p,
\end{equation}
where G is Newton's constant, $E_0=2 (m_p/M) G M_{\star} /R_{\star} $ is the maximum energy per branon mass that can lead to
capture,  $\rho_{dm}$ is the local dark matter density (for isolated white dwarfs) taken to be $\rho_{dm}=0.3 GeV/cm^3$. This value for $\rho_{dm}$ is conservative since current observations suggest
$\rho_{dm} \simeq 0.38 GeV/cm^3$, while some others indicate a value two times larger (see \cite{ilidio1,ilidio2} for details). Finally the probability p is given by $p=0.89 \sigma/\sigma_{cr}$, where the critical cross section is given by
\begin{equation}
\sigma_{cr} = 4 pb \left( \frac{R_{\star} }{R_{\odot} } \right)^2 \left( \frac{M_{\star} }{M_{\odot} } \right)^{-1}=4 \times 10^{-40} cm^2,
\end{equation}
while p saturates to unity if $\sigma > \sigma_{cr}$. Finally, the total branon mass accreted during the lifetime of the white dwarf $t_* \sim 1 Gyr$
is given by $M_{acc}=M N_{acc}$, where the accumulated number of branons is determined by solving the rate equation \cite{Jungman:1995df}
\begin{equation}
\frac{dN_{acc}}{dt} = F - \frac{\langle \sigma v \rangle}{V_b} \: N_{acc}^2,
\end{equation}
where $V_b$ is the volume of the sphere in which branons are mostly concentrated, and $\langle \sigma v \rangle \simeq 3 \times 10^{-26} cm^3/s$ is the branon
annihilation cross section required to reproduce the observed DM abundance \cite{planck2015}. Note that this a condition that $f,M$ must satisfy, and therefore only one of the two is independent. Choosing the branon mass to be the independent parameter, the brane tension factor as a function of $M$ can be seen in Fig. 1.
With the initial condition $N_{acc}(0)=0$, the rate equation
can be easily integrated, and thus the number of branons accumulated inside the star during its lifetime is given by
\begin{equation}
N_{acc} = \sqrt{\frac{F V_b}{\langle \sigma v \rangle}} 
\: \tanh{\left( \sqrt{\frac{F \langle \sigma v \rangle}{V_b}} \: t_*\right)}.
\end{equation}
It is worth mentioned that the exact solution above acquires a simpler form in two limiting cases, namely when the argument of the function $tanh(x)$ is very small
(or else when the annihilation cross section can be neglected), $x \ll 1$, and also when it is large (or else when after sufficiently long time branons reach the equilibrium. Then the two competing mechanisms in the rate equation cancel one another and the number of DM particles remain the same), $x \gg 1$. In the first 
case one finds
$N_{acc} \simeq F t_*$, which can be obtained from the rate equation neglecting the annihilation term, while in the second case one finds
\begin{equation}
N_{acc} \simeq  \sqrt{\frac{F V_b}{\langle \sigma v \rangle}},
\end{equation}
which can be obtained from the rate equation setting $dN/dt=0$. It is easy to verify that in our work, given the numerical values at hand, we can use the
previous formula for $N_{acc}$ at equilibrium.

For a gravitational collapse to take place inside the star the following three conditions have to be satisfied:

- First, in a system of non-interacting bosons only the uncertainty principle opposes the collapse, and the number of particles required for the gravitational collapse to occur is given by $N=(2/\pi) (M_p/M)^2$ \cite{kouvaris2}, and therefore the minimum mass of a self-gravitating lump that can form a black hole is $M_{cr}=M N=(2/\pi) (M_p^2/M)$. Thus, the first condition to be satisfied is
\begin{equation}
M_{acc} > M_{cr}.
\end{equation}

- The second condition comes from the fact that the newly-formed black hole must not emit Hawking radiation \cite{HR1,HR2} too fast. In fact, in the BH mass rate the Bondi accretion term \cite{bondi} must dominate over the energy loss due to the Hawking radiation \cite{kouvaris2}
\begin{equation}
\frac{4 \pi \rho G^2 M_{acc}^2}{c_s^3} > \frac{1}{15360 \pi G^2 M_{acc}^2},
\end{equation}
with $c_s$ being the speed of sound. Assuming a polytropic equation of state for a non-relativistic electron gas $P(\rho)=K \rho^{5/3}$ the speed of sound $c_s^2 = dP/d\rho$ is computed to be $c_s=\sqrt{(5 P)/(3 \rho)} \simeq 0.01$. This implies that $M_{acc} > M_2$, with $M_2$ being
\begin{equation}
M_2 = \left( \frac{c_s^3 M_p^8}{4 \pi^2 \rho \times 15360} \right)^{1/4} = 6.6 \times 10^{37} GeV.
\end{equation}

- Finally, the last condition comes from the onset of branon self-gravitation.
When the total branon mass captured inside a sphere of radius $r_*$ exceeds the mass of the ordinary matter within the same radius
\begin{equation}
M_{acc} > \frac{4 \pi \rho r_*^3}{3},
\end{equation}
the self-gravitation of branons dominate over that of the star \cite{kouvaris2}. Naively it is expected that most of
the branons are concentrated inside a radius $r_{th}$ given by \cite{kouvaris1}
\begin{equation}
r_{th}=\left( \frac{9 T}{8 \pi G M \rho} \right)^{1/2}.
\end{equation}
However, as first pointed out by Bose \cite{bose} and later expanded by Einstein \cite{einstein1,einstein2}, in a quantum gas made of bosons
the indistinguishability of the particles requires a new statistical description, now known as Bose-Einstein statistics.
If the temperature of the gas is low enough or the number density of particles is large enough, a new exotic form of matter
is formed. The Bose-Einstein Condensate (BEC) is driven purely by the quantum statistics of the bosons, and not by the interactions between them. The critical temperature is given by \cite{thesis}
\begin{equation}
T_c = \frac{2 \pi \hbar^2}{M k_B} \left( \frac{n}{\zeta(3/2)} \right)^{2/3} \simeq 3.3 \frac{n^{2/3}}{M},
\end{equation}
in our natural units, where $\zeta(3/2) \simeq 2.612$ is Riemann's zeta function, and $n$ is the number density of bosons,
i.e. branon DM particles. The BEC, considered to be the fifth state of matter after gases, liquids, solids and plasma, is manifested in the classical
example of the Helium-4 superfluidity \cite{superfluid}, and led to the Nobel Prize in Physics in 2001 \cite{nobel}.
The size of the condensed state is determined by the radius of the wave function of the branon ground state in the gravitational
potential of the star \cite{kouvaris2}
\begin{equation}
r_c = \left( \frac{8 \pi G \rho M^2}{3} \right)^{-1/4}.
\end{equation}


\section{Constraints on the branon DM parameter space}

First we employ the thermalization
condition $t_2 < t_0$ derived and used in \cite{kouvaris1}, with $t_2$ given by
\begin{equation}
\small
t_2 = 4 yr \left( \frac{M}{TeV} \right)^{3/2} \left( \frac{10^8 g\;cm^{-3}}{\rho} \right) \left( \frac{10^{-43} cm^2}{\sigma} \right)
\left( \frac{10^7 K}{T} \right)^{1/2},
\end{equation}
where T is the temperature and $\rho$ the matter density of the star, while $\sigma$ is the total S.I. branon-nucleus scattering
cross section. The thermalization condition implies a lower limit for the branon-nucleon cross section
\begin{equation}
\small
\sigma_n > \frac{4}{A^2} \left( \frac{M}{TeV} \right)^{3/2} \left( \frac{10^8 g/cm^3}{\rho} \right) 10^{-52} cm^2
\left( \frac{10^7 K}{T} \right)^{1/2},
\end{equation}
with $A=16$ since we have assumed a WD core consisting of $O^{16}$. Additionally for reference we also compute
$\sigma_n$ for Helium ($A=4$) and Carbon ($A=12$). Furthermore, the BEC is formed below the critical
temperature, $T < T_c$, so the condition for its formation is set by
\begin{equation}
\frac{3 N_{acc}}{4 \pi r_c^3} > \left(\frac{M T}{3.3} \right)^{3/2}.
\end{equation}
Our main results are summarized in Fig. 1 to 4.

First of all, given the conditions presented in the discussion above it is easy to verify that:

a) Whether a BEC is formed or not depends on the branon mass. As it can be seen in Fig. 2, when the branon mass is larger than $10 GeV$ the BEC formation requires
a branon-nucleon cross section so large that contradicts the limits from direct detection searches. On the other hand, for light branons the formation of a BEC
is possible.

b) When branons are not very light, $M > 2 keV$, $r_{th}$ is lower than the radius of the star, so branons are indeed
trapped inside the white dwarf. In addition, $r_c$ is lower than $r_{th}$ which implies that branons are indeed concentrated inside a sphere with radius $r_c$ and not inside a sphere with radius $r_{th}$ as it is expected if the BEC is not formed.

c) When branons are light, $M < 1 GeV$, the strongest condition for the black hole formation comes from the uncertainty principle, namely $M_{acc} > M_{cr}$, otherwise the condition becomes $M_{acc} > (4 \pi \rho r_{th}^3)/3$ (onset of branon self-gravitation).

Fig. 3 shows the allowed parameter space on the $\sigma_n-M$ plane for a branon mass $M > 10 GeV$ and for $A=16$. For a given branon mass, the branon-nucleon cross section must lie below the solid curve. For comparison we also show in the same plot the limits from direct detection experiments (dashed curve). Thus, for heavy branons white dwarfs fail to provide us with bounds better than current limits from DM direct detection searches. 
Furthermore, Fig. 4 corresponds to light branons with a mass $2 keV < M < 1 GeV$ and for $A=4,12,16$ from top to bottom. The black curve in the same figure
corresponds to the branon-nucleon cross section computed in the framework of the model (eq. (6)) taking into account the condition $\langle \sigma v \rangle \simeq 3 \times 10^{-26} cm^3/s$. Therefore, here too white dwarfs fail to provide us with bounds better than what the dark matter constrain can already tell us.

\section{Conclusions}

In the present article we have used for the first time white dwarf stars to constrain the parameter space of branon
dark matter. This new class of DM candidates are very well motivated within the framework of superstring theory.
It is known that superstring theory, the only consistent theory of quantum gravity, contains extended objects called branes,
and the brane-world proposal has been used, among other things to address the hierarchy problem of particle physics.
If the brane on which we live is flexible, the only relevant degrees of freedom are the Standard Model fields and the branons,
which are new scalar fields related to the brane fluctuations. Since they are stable, massive and weakly coupled
are natural DM candidates. Indeed it has been shown that branon DM particles are excellent dark matter candidates.
The parameter space is simple and consists of two mass scales only, namely
the branon mass M and the brane tension factor $f$. Given that WD do exist we were able to constrain the branon
parameter space. Our findings indicate that i) for heavy branons with a mass $M > 10 GeV$ white dwarfs fail to provide us with bounds better than current limits 
from DM direct detection searches, and ii) for light branons in the mass range $2 keV < M < 1 GeV$, which cannot be probed neither with current dark 
matter experiments nor with the next generation of detectors, the dark matter abundance constrain determines
$f$ as a function of $M$ in the range $0.1 GeV < M < 1 GeV$ for the branon mass and $1 GeV < f < 5 GeV$ for the brane tension factor.
Furthermore, our numerical results show that the limits from white dwarfs are not stronger than the dark matter abundance constrain.
Although this analysis is non-competitive with collider searches, it serves as a new and independent test.

\medskip


\begin{acknowledgments}
We wish to thank the anonymous reviewer for helping us to improve the quality of our manuscript.
The authors thank the Funda\c c\~ao para a Ci\^encia e Tecnologia (FCT), Portugal, for the financial support to the 
Multidisciplinary Center for Astrophysics (CENTRA),  Instituto Superior T\'ecnico,  Universidade de Lisboa,  through 
the Grant No. UID/FIS/00099/2013.
\end{acknowledgments}


\newpage




\end{document}